
\input phyzzx

\def\winf{$W^+_{1+\infty}$}
\def\win{W^+_{1+\infty}}
\def\woneplus{$W^+_{1+\infty}$}
\def\del{\partial}
\def\ba{{\bar A}}
\def\d{{\cal D}}
\def\tr{{\rm tr}}

\titlepage
\hfill{IASSNS-HEP-91/72}\break
\indent\hfill{TIFR-TH-91/51}\break
\title{Bosonization of Nonrelativistic Fermions and $W$-infinity Algebra}
\author{Sumit R. Das, Avinash Dhar, Gautam Mandal\foot
{e-mail: das@tifrvax.bitnet, adhar@tifrvax.bitnet,
mandal@tifrvax.bitnet.}}
\address{ Tata Institute of Fundamental Research, Homi Bhabha Road, Bombay
400 005, India}
\andauthor{Spenta R. Wadia
\foot{Supported by DOE grant DE-FG02-90ER40542.}
\foot{e-mail: wadia@iassns.bitnet.}
\foot{On leave from Tata Institute of Fundamental Research,
Bombay 400 005, India.}}
\address{School of Natural Sciences, Institute for Advanced Study,
Princeton, NJ 08540, U.S.A}
\abstract{We discuss the bosonization of non-relativistic fermions in one
space dimension in terms of bilocal operators which are naturally related
to the generators of $W$-infinity
algebra. The resulting system is analogous to the
problem of a spin in a magnetic field for the group $W$-infinity. The new
dynamical variables turn out to be $W$-infinity group elements valued in
the coset $W$-infinity/$H$ where $H$ is a  Cartan subalgebra. A classical
action with an $H$ gauge invariance is presented. This action is
three-dimensional. It turns out to be similiar to the action that describes
the colour degrees of freedom of a Yang-Mills particle in a fixed external
field.  We also discuss the relation of this action with the one we
recently arrived at in the Euclidean continuation of the theory using
different coordinates.}
\vfill
\endpage

\Ref\TOMONAGA{S. Tomanaga, Prog. Theor. Phys., 5(1950)544.}
\Ref\LIEB{See {\sl Mathematical Physics in One Dimension \sl},
Ed. E.H. Lieb and D.C. Mattis (Acdemic Press, 1966).}
\Ref\BIPZ{E. Brezin, C. Itzykson, G. Parisi and J. Zuber, Comm. Math.
Phys. 59 (1978) 35.}
\Ref\JS{A. Jevicki and B. Sakita, Nucl. Phys. B 165 (1980) 511.}
\Ref\DJEV{S.R. Das and A. Jevicki, Mod. Phys. Lett. A5 (1990) 1639}
\Ref\SW{A.M. Sengupta and S.R. Wadia,
Int. J. Mod. Phys. A6 (1991) 1961.}
\Ref\GK{D. Gross and I. Klebanov,
Nucl. Phys. B352 (1990) 671.}
\Ref\MSWA{G. Mandal, A. Sengupta and S.R. Wadia,
Mod. Phys. Lett. A6 (1991) 1465.}
\Ref\MORE{G. Moore, Rutgers Preprint RU-91-12
(1991).}
\Ref\POLC{J. Polchinski,
Texas Preprint UTTG-16-91 (1991).}
\Ref\KS{D. Karabali and B. Sakita, City College Preprint,
CCNY-HEP-91/2.}
\Ref\DDMW{S.R. Das, A. Dhar, G. Mandal and S.R. Wadia, ETH, IAS and
Tata preprint, ETH-TH-91/30, IASSNS-HEP-91/52 and TIFR-TH-91/44.}
\Ref\WADIA{S.R. Wadia, Seminar given at Syracuse University, 8 April,
1991.}
\Ref\AJ{J. Avan and A. Jevicki, Brown Preprints BROWN-HET-801,824 (1991).}
\Ref\POLCHINSKI{D. Minic, J. Polchinski and Z. Yang,
Texas preprint, UTTG-16-91.}
\Ref\BAKAS{I. Bakas, in {\sl Supermembranes and Physics in $2+1$
Dimensions}, Ed. M. Duff {\sl et al} (World Scientific, 1990).}
\Ref\POPE{C. N. Pope, L.J. Romans and X. Shen, in {\sl Strings 90}, Ed. R.
Arnowitt {\sl et al} (World Scientific, 1991).}
\Ref\WITTEN{E. Witten, IAS preprint, IASSNS-HEP-91/51.}
\Ref\POLYAKOV{A.M. Ployakov and I. Klebanov,
Princeton University preprint.}
\Ref\BERGSHOEFF{E. Bergshoeff, P.S. Howe, C.N. Pope, E. Sezgin and K.S.
Stelle, Texas A\& M, Imperial and Stony Brook preprint, CTP TAMU-25/91,
Imperial/TP/90-91/20  and  ITP-SB-91-17.}
\Ref\MOORE{G. Moore and N. Seiberg, Rutgers and Yale preprint, RU-91-29 and
YCTP-P19-91.}
\Ref\WIGNER{E.P. Wigner, Phys. Rev. 40(1932) 749.}
\Ref\BAL{A.P. Balachandran, G. Marmo, B.S. Skagerstam and A. Stern,
{\sl Gauge Symmetries and Fiber Bundles},
Springer Lecture notes in Physics vol. 188.}
\Ref\JW{S. Jain and S.R. Wadia, Nucl. Phys. B258 (1985) 713.}
\Ref\BLR{A.P. Balachandran, F. Lizzi, V.G.J. Rodgers and A. Stern, Nucl.
Phys. B 256 (1985) 525.}
\Ref\SHAPERE{See {\sl Geometric Phases in Physics},
Ed. A. Shapere and F. Wilczek (World Scientific, 1989).}

\underbar{Introduction}:

In this paper we discuss the  bosonization of non-relativistic
fermions in one space dimension. This is a very old subject beginning with
the classic work of Tomonaga [\TOMONAGA] in which he succeeded in deriving
the equations of motion of the collective excitation (sound wave) near the
fermi surface, under certain approximations. There have been many
subsequent studies of this problem [\LIEB].  More recently there was a
revival of interest in this issue in the discussions of the $c=1,d=2$
string theory. As is well known, this model is formulated as a many fermion
problem in a background potential [\BIPZ].  The bosonization of this
theory has been discussed by many authors from different viewpoints
[\JS-\KS].  The common feature of all these formulations is that the basic
boson variable is chosen to be local and is essentially the bilinear
$\psi^\dagger(x,t)\psi(x,t)$. In a recent work [\DDMW] we had argued that
a more natural set of variables for this problem
are the generators of the \winf algebra which are in one-to-one
correspondence with the ``bilocal" operator $\psi(x,t)\psi^\dagger(y,t) \equiv
\phi(x,y,t)$.  This non-local description is also satisfactory from the
point of view that this model describes a string theory.
In the present problem
not all of $\Phi(t)$
\footnote*{Here and in the following
we shall adopt the matrix notation of [\DDMW]
and not indicate the $x,y$-dependence explicitly, unless desirable for
clarity.  Thus $\phi(x,y,t)$ are
components of the matrix $\Phi(t)$.}
constitute an independent complete set of
variables. We have shown [\DDMW] from an analysis of the classical limit
that the independent variables belong to the coset $G/H$ where
$G$ is the \winf-group (group of area-preserving diffeomorphisms in two
dimensions) and $H$ is a Cartan subgroup. As we emphasized in [\DDMW], the
coset $G/H$ constitutes the `phase space' of the theory. We also presented
a phase space action in terms of a certain parametrization of the coset
variables (see equation (9.18) of ref [\DDMW]).
In the present paper we employ a different parametrization of the coset to
discuss the bosonization
and derive the classical
action in terms of these variables.
This constitutes a solution of
the bosonization problem.
We hope that besides the solution of the bosonization problem
that we have obtained here this formulation will help resolve
some of the perplexing questions of the $d=2$ string theory, especially
those related to the puzzling differences in the Minkowski and Euclidean
formulations and of course the issue of black holes.

It is appropriate at this point to say a few words about the recent
developments in this subject. The conservation laws of the fermionic model
were first presented in [\SW,\GK]. Their local form and the algebra of the
currents was obtained by the present authors [\WADIA,\DDMW] and in [\AJ,
\POLCHINSKI]. The possible role of the algebra of area-preserving
diffeomorphisms and higher spin states in two dimensions was pointed out
in [\BAKAS, \POPE]. A precise connection of this algebra with the vertex
operators of the $d=2$ string theory was discussed in [\WITTEN,\POLYAKOV].
The role of the $W$-infinity algebra in the matrix model was discussed in
[\DDMW] and from a different point of view by [\AJ, \MOORE].

\underbar{Gauge theory formulation of the $c=1$ matrix model}:

In the gauge theory formulation [\DDMW] of the double scaled fermion
field theory  the action is given by
$$S[\Psi, \Psi^\dagger,\ba]=\int dt\; \Psi^\dagger(t) (i\del_t+ \ba(t))\Psi(t)
\eqn\one$$
where $\ba(t)$ is a background gauge field and we have used the notation
of [\DDMW]. For the double scaled $c=1$ matrix model
$\ba_{xy}(t)=1/2\; (\del_x^2+ x^2)\delta(x-y)$.
This action has the background gauge invariance
$$\eqalign{
\Psi(t) &\to V(t) \Psi(t) \cr
\ba(t)  &\to  V(t)\ba V^\dagger(t) + i V(t) \del_t V^\dagger(t) \cr
} \eqn\onea $$
where $V(t)$ is a unitary operator on the single-particle Hilbert space in
which $\Psi(t)$ is viewed as a vector
\footnote*{A more appropriate notation might be to write the ket
$|\Psi(t)>$, which is equivalent to thinking of the second-quantized field
as a many-body wave-function.  In this notation $\Phi(t)$ may be written
as $|\Psi(t)> <\Psi(t)|$. The components $\phi(x,y,t)$ may then be written
as $<x|\Psi(t)><\Psi(t)|y>=\psi(x,t)\psi^\dagger(y,t)$.}
with components $\psi(x,t)$. In ref [\DDMW] we identified the
generators $W^{(r,s)}$ of the symmetry algebra \woneplus\ of the theory.
We also pointed out that the relevenace of the symmetry algebra for this
problem lies in the fact that the generators $W^{(r,s)}$ are themselves
the basic (boson) dynamical variables. Their definition involved
polynomials of the coordinate and momentum of a single fermion. Towards
the end of ref [\DDMW], in an added note, we also introduced a
generating function for the operators $W^{(r,s)}$ in a ``classical" phase
space $(p,q)$ by the formula
$$W(p,q,t)= 1/2 \int dx\; e^{ipx}\; \psi^\dagger(x+q/2,t)\; \psi(x-q/2,t)
\eqn\foura$$
The generators $W^{(r,s)}$'s are then given by
$$
W^{(r,s)}
= \int da \; db\; F_{rs}(a,b) \;
W({b+a\over 2},{b-a\over 2},t)
\eqn\fourdc $$
where
$$F_{rs}(a,b)= 2 \cos({ab\over 2})\; (i\del_a)^r (i\del_b)^s
\delta^{(2)}(a,b) \eqn\fournew$$
The functions $F_{rs}$ carry a representation of the $W$-infinity
algebra.

The operators $W(p,q,t)$ satisfy a simple algebra
$$ [W(p,q,t), W(p',q',t)]= i\;\sin ({\hbar\over2}(pq' - qp'))
\; W(p+p', q+q',t)\eqn\fouraa$$
In \fouraa\ we have explicitly exhibited the string coupling constant
$ g_{str}\sim\hbar $, so that in the limit $\hbar \to 0$ we regain the
structure constants of the algebra of area-preserving diffeomorphisms on
the plane.
The quantum deformation of the classical algebra of area-preserving
diffeomorphisms
has been discussed in [\BERGSHOEFF]. For the present problem it was
mentioned in [\POLYAKOV].

Equations \foura\ and \fouraa\ can be
understood in a framework that formulates
quantum mechanics in terms of a
classical phase space which is due to Wigner [\WIGNER]. To explain
this, let us introduce the Heisenberg-Weyl group with elements
$$g_{p,q}= e^{-i(q\hat P - p \hat X)} \eqn\fourab$$
where $(p,q)$ is a point in the plane ${\bf R}^2$ and $\hat P$ and $\hat
X$ satisfy the commutation relation $[\hat P, \hat X]= -i \hbar$.
The group multiplication law is given by
$$g_{p,q} g_{p',q'} = e^{i\hbar
(pq'-qp')/2} g_{p+p', q+q'} \eqn\fourea$$
Now noting that
$g_{p,q}= e^{-iq{\hat P}/2}
e^{ip{\hat X}} e^{-iq {\hat P}/2} $, we see that
$$W(p,q,t)=1/2 \int \;dx \;  \psi^\dagger(x,t)\;
g_{p,q} \; \psi(x,t) \eqn\fourd$$
One can now immediately recognize that
the algebra \fouraa\ is an immediate consequence of the group law
\fourea.

We now point out that the
time evolution of $W(p,q,t)$ is governed by a simple equation. In
ref [\DDMW] we had shown that the time evolution of $W^{(r,s)}(t)$
is given by the equation
$$ \del_t W^{(r,s)}(t) =\epsilon
 (r-s) W^{(r,s)}(t) \eqn\fourdi$$
with solution $W^{(r,s)}(t)= \exp(\epsilon(r-s)\;t)\; W^{(r,s)}(0)$
($\epsilon=1$ in Minkowski and $-i$ in Euclidean space). An equally simple
equation governs $W(p,q,t)$
$$[\del_t +\epsilon(p\del_q + q\del_p)] W(p,q,t)=0 \eqn\fourdj$$
We see that the time development of $W(p,q,t)$ is simply a $U(1)$-rotation
in the phase plane. The solution is
$$W(p,q,t)= \sum_{m=-\infty}^\infty e^{\epsilon m(\theta-t)}\;
W_m(\tau) \eqn\fourdk$$
where we have parametrized the phase plane by
$p= \exp(-\tau)\;\cosh(\epsilon\theta),\; q=\exp(-\tau)\;
\sinh(\epsilon\theta)$
In ref [\DDMW] we had identified $\tau$ with the Liouville
coordinate of the continuum string theory.

Equations \fouraa\ and \fourdj\ constitute an exact boson representation of
the non-relativistic fermion theory under consideration.

\underbar{Boson representation in terms of the bilocal operator}:

The discussion in this and the subsequent sections is best done in terms of
the original bilocal operator $\Phi(t)$.
The operator $W(p,q,t)$ is related to it by a fourier transform as in
\foura.
The first point to note is that
the equation of motion for $\Psi(t)$ following from \one\ implies the
following equation for the bilocal operator $\Phi(t)$:
$$i\del_t\Phi(t) + [\ba(t), \Phi(t)]=0 \eqn\two$$
Moreover, the equal time algebra for $\Phi(t)$ following from the
anticommutation relation of $\Psi(t)$ and $\Psi^\dagger(t)$ is
$$[\Phi^\alpha (t), \Phi^\beta(t)]= i\sum_\gamma C^{\alpha \beta}_\gamma
\Phi^\gamma(t)   \eqn\three$$
where $\alpha, \beta,\cdots$ denote the components $(x_\alpha, y_\alpha),
(x_\beta, y_\beta),\cdots$ and $C^{\alpha\beta}_\gamma$ are the structure
constants
$$C^{\alpha\beta}_\gamma = -i\delta(x_\beta-
y_\alpha)\delta(x_\gamma-x_\alpha )\delta(y_\gamma-y_\beta) +
i\delta(x_\alpha - y_\beta)\delta(x_\gamma- x_\beta) \delta
(y_\gamma - y_\alpha)    \eqn\four $$
The sum over $\gamma$ in \three\ denotes the integration over the
variables $(x_\gamma, y_\gamma)$.
The structure constants in \four\ are related to the
standard ones for \woneplus\ by a redefinition of the generators.

Equations \two\ and \three\
show that $\Phi(t)$  is like a spin variable
and the present problem is like that of a spin in a magnetic field.
The hamitonian for a quantum spin $S^i$ in a magnetic field $B_j$ is
$h=S^j B_j$ and the equation of motion is $i\del_t S^i = [h, S]^i=
iC^{ij}_k B_j S^k$ using  $[S^i, S^j]=iC^{ij}_k S^k$. If we define the
matrix of operators $S= S^i T_i$ and a matrix-valued magnetic field $B=
-B^j T_j$ then the equation of motion becomes $i\del_t S +[B,S]=0$ in exact
analogy with equation \two. In the usual case of the quantum spin the
$T_i$'s correspond to the generators of $SU(2)$; in our case the algebra
is  \woneplus.

It  immediately follows  from \two\ that the set of operators ${\rm tr}
(\Phi(t))^n, n=1,2,\cdots$ are constants of motion. In fact, these are
precisely the Casimir operators for the \woneplus\ algebra \three.
This is clear from the fact that $\tr (\Phi(t))^n$ is invariant under
$\Phi(t) \to V \Phi(t) V^{-1}$ where $V$ is any element of the
\winf-group. Since the above rotation is generated by the $\Phi^\alpha$'s
themselves,  this implies that
$$ [\tr (\Phi(t))^n, \Phi^\alpha]=0 \quad {\rm for \; all}\; \alpha
\eqn \fiveminus$$
It is
the existence of these Casimir invariants that reduces the the number of
independent variables of the problem.

The identification of $\Phi(t)$ with a spin-like variable and the
constraints coming from the Casimir invariants suggest the following
parametrization of the bilocal operator
$$\Phi(t)= U(t) \Lambda(t) U^{-1}(t) \eqn\five$$
where $U$ is an element of the \woneplus\ group and $\Lambda(t)$ is a
diagonal matrix. Since ${\rm tr}(\Phi(t))^n= {\rm tr} (\Lambda(t))^n,
n=1,2,\cdots$ are all fixed numbers, it follows that $\Lambda$ is a fixed
time-independent matrix, determined by the values of the Casimir
invariants. With this ingredient we
see that \two\ implies the following equation of  motion for $U(t)$
$$ [\Lambda(t), iU^{-1}\del_t U + U^{-1}\ba U]=0 \eqn\fiveaaaa$$
which just states that the constant matrix $\Lambda$ commutes with the
gauge orbit of $\ba$.
This equation of motion can  be
derived from the following
classical action by
varying $U$
$$S[U, \ba, \Lambda]= \int dt\; {\rm tr} \{(iU^{-1}\del_t U + U^{-1}
\ba U)\Lambda\} \eqn\six$$
A simliar action has previously appeared in the problem  of Yang-Mills
particle [\BAL]. The analogy is exact if the centre of mass motion
of the particle is
frozen because then only the colour degrees of freedom  are active.
It has also appeared previously in the discussions of quantization of
baryonic collective excitations around
the Skyrme soliton [\JW,\BLR]. In these
works it was shown that for the collective excitations the WZW term reduces
to the action \six\ with $\ba=0$. It is also
interesting to note that there is
a natural two-dimensional representation for this action [\JW,\BLR].

It is important to realize that the action \six\ is really a 3-dimensional
action. To emphasize this point we write out the r.h.s. of \six\ in
detail. Using the notation $U_{xy}(t) \equiv u(x,y,t),
\Lambda_{xy} \equiv \lambda(x)\;\delta(x-y)$, we have
$$S[U,\ba,\Lambda]=\int\;dt\;dx\;dy\;\lambda(y)\;u^*(x,y,t)\;[i\del_t+
\ba(x,\del_x)]\;u(x,y,t)  \eqn\sixy$$
where $u(x,y,t)$ satisfies the unitarity constraint
$$\int dx'\; u^*(x',x,t) u(x',y,t) = \int dx' u(x,x',t) u^*(y,x',t)
=\delta(x-y) \eqn\sexy$$
and  $\ba \equiv {1\over 2}(\del^2_x+x^2)$
is the background  specific to the $c=1$ matrix model.

An important aspect of the parametrization in \five\ is that $U(t)$ is
not uniquely determined. In fact, $U(t)$ and $U(t)g(t)$ such that
$g(t) \Lambda g^{-1}(t)= \Lambda$ are equivalent. For generic $\Lambda$,
the set of all such elements $g(t)$ forms a Cartan subgroup $H$ of
\woneplus. The physical degrees of freedom of this system, therefore,
belong to the coset $W^+_{1+\infty}/H$. One would therefore expect  the
action in \six\ to have the local gauge symmetry $U(t) \to U(t) g(t)$.
This indeed turns out to be correct [\BAL-\BLR].  We also note that \six\ has
the background gauge invariance of the action \one, i.e
\six\ is invariant under
$$U(t) \to V(t) U(t),\qquad\quad \ba(t) \to V(t)\ba(t) V^{-1}(t)+ iV(t)
\del_t V^{-1}(t)\eqn\seven $$

It is important to realize that the action  \six\
is first order in time derivative and therefore it
must be interpreted as a phase space action.
Furthermore, as we have discussed above, this action has a gauge
inviariance which restricts the physical degrees of freedom to the coset
$\win/H$. Thus the action \six\ is a phase space action with the phase
space being the coset. In [\DDMW] we arrived at an identical conclusion
from the analysis of the classical limit, and wrote down a bosonised
action in terms of a certain parametrization of the coset. As we
shall see later, for the simple example of $SU(2)/U(1)$ these two actions
turn out to be identical.

How does one determine the constant matrix $\Lambda$ appearing in \six?
As we shall see below, one can explicitly calculate
all the Casimir invariants in the fermion field theory described by the
action \one.  Since the knowledge of all the Casimirs is equivalent to the
knowledge of $\Lambda$, we can in this way determine $\Lambda$.

We have proved in [\DDMW] that the fermion Fock space in the zero fermion
number sector forms an irreducible representation of \winf algebra.
Therefore, to find out the values of the Casimirs it is sufficient to find
out their values on the fermi vacuum. Consider, for example, the first
Casimir $C_1= \tr\; \Phi(t)$. Expanding the fermion field in terms of a
complete orthonormal basis of energy eigenstates
$$ \eqalign{
\psi(x,t) &= \sum_n e^{-i E_n t} u_n(x) c_n \cr
\psi^\dagger (x,t) &= \sum_n e^{i E_n t} u_n^*(x) c^\dagger_n \cr
\{ c_n, c^\dagger_m\} &= \delta_{nm} \cr
} \eqn\sevena$$
we can write this Casimir as
$$C_1= \sum_n c_n c^\dagger_n \eqn\sevenb$$
It is clear that $C_1$ on any state measures the total number of
unoccupied levels. To find its numerical value it is sufficient
to find its value on the fermi vacuum.
Denoting the fermi level by $n=n_0, E_{n_0}=\mu$,
we see that  the
value of $C_1$ is
$$C_1= \sum_{n_0}^\infty 1 = \int_\mu^\infty dE\; \rho(E) \eqn\sevenc$$
where  $\rho(E)$ is the level density.
Similarly, the second Casimir $C_2= \tr\; (\Phi(t))^2 $ has the expression
in terms of the oscillators
$$C_2= \sum_{m,n} c_m c^\dagger_n c_n c^\dagger_m \eqn\sevend$$
This  can be rearranged and put in the form
$$C_2= C_1 (1+ \tilde {C_1}) \eqn\sevene$$
where $\tilde {C_1}$ denotes the total number of occupied levels
$$\tilde {C_1}= \sum_n c^\dagger_n c_n \eqn\sevenda$$
On the Fermi vacuum $\tilde {C_1}$ evaluates to
$$\tilde {C_1}= \sum_{-\infty} ^{n_0} 1 = \int_{-\infty}^{\mu} dE\;
\rho(E) \eqn\sevenf$$
In general, for the $n$-th Casimir $C_n = \tr\; (\Phi(t))^n$, we find
$$C_n = C_1 (1+ \tilde {C_1})^{n-1} \eqn\seveng$$
Equation \seveng\ is ill-defined as
it stands, since both $C_1$ and $\tilde
{C_1}$ diverge. It is, however, sufficient to introduce regulators at the
two ends of the spectrum to regularize all the Casimirs.
The dependence on these regulators should drop out of all
physical calculations in any well-defined renormalization scheme.
\vfill
\eject

\underbar{Path integral and effective boson field theory}:

So far we have argued in the operator formulation
that the action \six\ is the correct
bosonization of the problem. We shall now present arguments
within the fermionic functional integral for the action \one\
to arrive at the same conclusion.

In the fermion field theory described by the action \one, we are in
general interested in an arbitrary correlation function of the fermion
bilinears $\psi(x,t) \psi^\dagger(y,t)$:
$$\eqalign{
G(x_1,y_1,t_1;\cdots;x_n,y_n,t_n)&\;\cr
={1\over Z[\ba]}
\int \d\Psi\d\Psi^\dagger\; \psi(x_1,t_1)\psi^\dagger(y_1,t_1)&
\cdots\psi(x_n,t_n)\psi^\dagger(y_n,t_n) \exp(iS[\Psi,\Psi^\dagger,\ba])
\cr}
\eqn\sevenh$$
where $Z[\ba]$ is the parition function for the action \one.
A natural boson representation for this correlation function would be the
one in terms of the bilocal operator $\phi(x,y,t)$. Clearly the classical
action for $\phi(x,y,t)$ should be such that the correlation function of
$\phi(x_1,y_1,t_1),\cdots,\phi(x_n,y_n,t_n)$ exactly reproduces \sevenh.
One may think of this boson action as an effective action for the physical
excitations of the theory in complete analogy with the mesonic effective
actions of QCD. Moreover, as we argued in [\DDMW], since one can create
all the physical states of the theory by the action of an arbitrary string
of the bilocal operators on the fermi vacuum, such a classical bosonic
action would represent a complete bosonization of the theory.

Our strategy for the deriving this boson action is to
insert the following identity  in the fermion field theory partition
function which enforces the defintion of the bilocal operator
$\phi(x,y,t)$ in terms of the fermion fields
$$1= \int {\cal
D}\Phi\;\delta[\phi(x,y,t)-\psi(x,t)\psi^\dagger(y,t)]
\eqn\eight$$
Using the exponential representation for the delta-functional
$$\delta[\phi(x,y,t)-\psi(x,t)\psi^\dagger(y,t)]= \int {\cal D}A \exp
(i\int dt\; {\rm tr} A(t)\Phi(t)) \exp (i\int dt\; \Psi^\dagger (t) A(t)
\Psi(t))\eqn\nine $$
the partition function  can be written   as
$$Z[\ba]=\int\d\Phi\d A \exp(i\int dt\; \tr A(t)\Phi(t)) Z[A+\ba] \eqn\ten$$
where
$$ Z[A]= \int \d\Psi^\dagger \d
\Psi \exp(i\int dt\;\Psi^\dagger(t)(i\del_t+ A(t))\Psi(t)
\eqn\tena$$
Clearly, $A$ is like a source for the fermion bilocal operator
in \tena, and
$\log\; Z[A+\ba]$ is the generating functional of connected correlation
functions for these in the background $\ba$.
The integration over $A$ in \ten\ performs an inverse
fourier transform of $Z[A+\ba]$ and hence gives us the desired classical
bosonic action.

Let us now make the following change of
variables in the functional integral
$$\Phi(t)= U(t) \Lambda(t) U(t)^{-1}, A(t)\to U(t)A(t)U(t)^{-1},
\Psi(t) \to U(t)\Psi(t) \eqn\eleven$$
where $\Lambda(t)$ is diagonal.
The partition function now becomes
$$ Z[\ba]=\int\d U\d\Lambda\;J(\Lambda)
\d A \exp(i\int dt \;\tr A(t)\Lambda(t)) Z[A+\ba^U] \eqn\twelve$$
where
$$\ba^U= iU^{-1}\del_t U + U^{-1}\ba U \eqn\twel$$
and $J(\Lambda)$ is the van der
Monde Jacobian coming from the change of variables from the hermitian
matrix $\Phi$ to its eigenvalues $\Lambda$ and ``angles'' $U$.
We do not expect any anomalies in the fermionic measure because
we are dealing with non-relativistic fermions.

To be more precise the integration over $U$ should really be over a
coset  as is clear from the
counting of the degrees of freedom of $\Phi$ on the one hand (remember it
is a hermitian matrix) and $\Lambda$ and $U$ on the other hand.
But, as we have mentioned above, the action
of the  $U$ variables turns out to be \six\ which
has a gauge symmetry which enables
us to gauge away the extra degrees of freedom. Therefore the
integration over the $U$'s can be extended over the whole group.

We emphasize
that the measure for
$U$-integration is the usual Haar measure  and that there is no
$U$-dependent contribution from the Jacobian. As a result the entire
classical action for $U$ comes from the coupling of the dummy gauge field
$A(t)$ to the eigenvalues $\Lambda$. In fact, this action, obtained by
simply  shifting in \twelve\ the integration variable $A$ by $\ba^U$,
is identical to \six\ provided we can show that $\Lambda$
is independent of time. We have seen that this is true in
the operator formalism where we argued
that $\Lambda$ is determined in terms of the Casimirs
which are time-independent numbers. In the functional integral,
integration over the dummy gauge field $A$ and
the fermions gives rise to the delta functional $\delta[\del_t\Lambda]$.
Let us explain in some detail how that happens.

Since $\Lambda$ is diagonal it couples only to the diagonal
elements of $A$. So the only source of dependence on $\Lambda$ in the $A$
-$\Psi$
integration in \twelve\ is that over these diagonal elements $A_{xx}(t)$ of
$A$. This problem is like QED with the gauge field coupled to an external
source $J_\mu$ and without the kinetic term for the gauge field.
If the external
source has a longitudinal piece then integration over the gauge degrees of
freedom of the gauge field does not give rise to the gauge volume but
to a delta functional of $\del_\mu J_\mu$. In the present case, we can
decompose $A_{xx}(t)$ into gauge and physical degrees of freedom as
$$A_{xx}(t) = \del_t\epsilon(x,t) + a(x)\eqn\thirteen$$
The $A$-$\Psi$ functional integral in \twelve\ can then be written as
$$\int \d \epsilon(x,t) \int \d a(x) \;\exp(i \int dt\; dx\; (a(x)
\Lambda(x,t)  - \epsilon(x,t) \del_t \Lambda(x,t)) \; Z[a,\epsilon]
\eqn\fourteen$$
where
$$Z[a,\epsilon] = \int \d A \int \d \Psi^\dagger\; \d \Psi \exp(i \int dt\;
\Psi^\dagger(t) ( i\del_t + \del_t \epsilon(t)+ a + A(t)) \Psi(t))
\eqn\fifteen
$$
In \fifteen\ $\epsilon(t)$ and $a$ are diagonal matrices with components
$\epsilon(x,t)$ and $a(x)$,
and the gauge field $A(t)$ has only off-diagonal elements
non-zero which are being integrated over. Since $\epsilon(t)$ is diagonal
we can remove it from the r.h.s. of \fifteen\ by the change of variables
$$\eqalign
{ A(t) & \to e^{i\epsilon(t)} A(t) e^{-i \epsilon(t)}\cr
\Psi(t) & \to e^{i \epsilon(t)} \Psi(t) \cr
} \eqn\sixteen$$
which does not affect $a$ since it is also diagonal. In effect this means
that $Z[a,\epsilon] = Z[a]$ is actually independent of $\epsilon$. The
$\epsilon$ integration in \fourteen\ is, therefore, trivial and gives rise
to the delta-functional $\delta[ \del_t \Lambda(x,t)]$ as stated above.

We now come to the
integration over the constant matrices $\Lambda$ in
\twelve. That there is an integration over $\Lambda$
may seem surprising in view
of our demonstration earlier that
$\Lambda$ is a fixed $c$-number matrix determined by the values of
Casimirs. What this means is that in the functional integral there must be
delta-functions which enforce this. We must remember that the classical
action \six\ is a phase space action and therefore physical states are
functionals of only ``half"  the coset variables (the coordinates). Our
demonstration that the Casimirs have fixed values on such wave-functionals
implies that these delta-function constraints arise in the functional
integral only after integrating out the other "half" of the variables
(the momenta).

We may therefore write the partition function in \twelve\ as
$$Z[\ba]=\int \d\Lambda\; J(\Lambda)\; \tilde Z[\Lambda] \; \int \d U
e^{-iS[U,\ba,\Lambda]} \eqn\newtwelve$$
where
$$\tilde Z[\Lambda]= \int \; \d A e^{i\int dt\; \tr\; (A(t)\Lambda(t))}
Z[A] \eqn\newtwelvea$$
and it is understood that the $\Lambda$-integration will eventually get
frozen. We can similarly  argue that the boson representation for the
correlation
function in \sevenh\ is
$$\eqalign{
G(x_1,y_1,t_1;\cdots;x_n,y_n,t_n) =&
{1\over Z[\ba]} \int \d\Lambda\; J(\Lambda)\; \tilde Z[\Lambda] \cr
\int & \d U\; (U(t_1)\Lambda U^{-1}(t_1))_{x_1y_1} \cdots
(U(t_n)\Lambda U^{-1}(t_n))_{x_ny_n} e^{-iS[U, \ba, \Lambda]}\cr}
\eqn\newtwelveb$$
Since the $\Lambda$-integration in both the numerator and the denominator
in \newtwelveb\ is
eventually frozen, we may effectively write the above as
$$
G(x_1,y_1,t_1;\cdots;x_n,y_n,t_n) =
{\int \d U\; (U(t_1)\Lambda U^{-1}(t_1))_{x_1y_1} \cdots
(U(t_n)\Lambda U^{-1}(t_n))_{x_ny_n} e^{-iS[U, \ba, \Lambda]}
\over
\int \d U\; e^{-iS[U, \ba, \Lambda]} }
\eqn\newtwelvec$$
This represents a complete bosonization of the theory  and we have thus
derived the effective classical bosonic action \six\ in the functional
formulation.
\vfill
\eject

\underbar{The example of $SU(2)/U(1)$}

Finally, an important question to ask is how to paramatrize $U$ in
terms of physical coordinates and momenta. This requires fixing a gauge to
restrict $U$ to the appropriate coset and choosing a polarization.
A simple
illustrative example is that of $SU(2)/U(1)$
[\BAL, \SHAPERE].
For the simple choice of $\Lambda= \ba= \sigma_3$, the action \six\ in
this case becomes
$$ S = \int dt\; \tr\; (iU^{-1} \del_t U \sigma_3 \; + U^{-1} \sigma_3
U \sigma_3) \eqn\sixteena $$
where $U(t)$ are $SU(2)$ group elements.
It is easy to check that this action has the local gauge symmetry $U(t)
\to U(t) \exp(i \theta(t) \sigma_3) $. Therefore the phase space in this
case is the manifold $SU(2)/ U(1) = S^2$.
There is a natural action of $SU(2)$ on $S^2$
if we think of $S^2$ as $CP^1$. In other words let us represent points on
the sphere by a pair of complex numbers $(z_1, z_2)$ with the
understanding that $(z_1, z_2)$ and $(\lambda z_1, \lambda z_2)$ represent
the same point ($\lambda$ is any non-zero complex number).
Except when $z_2=0$ we can describe the point $(z_1, z_2)$ by the
``inhomogeneous coordinate" $z= z_1/ z_2$. Now an $SU(2)$ matrix acts on
the pair $(z_1, z_2)$ by sending it to $(az_1+ bz_2, cz_1, dz_2),\;
ad-bc=1$. This implies that
$$U: z \to {az+b \over cz+d },\;\;ad -bc =1 \eqn\seventeen$$
We would now like to parametrize points of the sphere by group elements in
the following fashion. Let us fix a point on the sphere, say $(z_1, z_2)
=(0,1)$ or equivalently $z=0$ (we can think of this point
as the North pole if we identify this $z$ with the stereographic
coordinate). Let us ask the question which group element $U$ takes us
from the North pole to a given point $z$.
Of course the answer is not unique
because any digaonal matrix ($b=c=0$) will keep $z=0$ at $z=0$,
according to \seventeen. This is simply a reflection of the fact that the
2-sphere is a coset of $SU(2)$. Let us
choose the following representative
element
$$
U(z)= \sqrt{1\over 1+ z\bar z}
\left(\matrix{1 & z\cr
       -\bar z &1 \cr}\right)$$
Clearly this sends $z=0$ to $z$.

Once we make this gauge choice it is simple to work out the action
\sixteena\ in terms
of $z, \bar z$.
$$S = 2\int dt \;{{i\over 2}(z \del_t
\bar z-\bar z\del_t z) +1-2\bar z z \over 1+ z\bar z}
\eqn\sixteenb$$
In terms of the variable $w= z/\sqrt{1+ z\bar z}$ the action becomes
$$S = 2\int dt\; [{i \over 2} (w\del_t\bar w- \bar w \del_t w)+ 1 -
3 w\bar w ]\eqn\sixteenc$$
This may be compared with the action we had written down in [\DDMW] (see
eqation (9.18) and the footnote following it). These two actions agree,
apart from a trivial shift of the hamiltonian by a constant
number and a rescaling of time.

\underbar{Concluding Remarks}

In summary, we have presented a classical action for the bosonic
excitations of the $c=1$ fermion field theory. The action is a phase space
action written in terms of  \woneplus\ group element $U(t)$ valued in a
coset $\win/H$. The action is naturally written in three dimensions.
An important open problem that remains is to find a
parametrization of the coset that lends itself to a two-dimensional
spacetime interpretation of the spectrum of the model. To this end, it
might be appropriate to use "mixed" bases such that
$u_n(x,t) = <x| U(t)|n>,\; u_n^*(x,t)=<n| U^\dagger(t)|x>$, where $|n>$
are the eigenstates of $\ba,\; \ba|n>= E_n|n>$.
Taking $\Lambda$ to be diagonal in the $|n>$ basis, i.e. $\Lambda=
\sum_n \lambda_n |n> <n|$, the  action reads in the mixed bases
$$S= \sum_n \lambda_n\int dt\;dx\;
u^*_n(x,t) [i\del_t+ \ba(x, \del_x)] u_n(x,t)\eqn\seventeen$$
The unitarity constraints on the $u_n(x,t)$ are
$$\eqalign{
\sum_n u_n^*(x,t) u_n(y,t)=& \delta(x-y)\cr
\int dx\; u_n^*(x,t) u_m(x,t) =& \delta_{nm}\cr}
\eqn\eighteen$$
Using \eighteen\ one can check that for generic $\lambda_n$
\seventeen\ has the gauge invariance $u_n(x,t)
\to \exp(i\theta_n (t)) u_n(x,t)$.
For non-generic $\lambda_n$ (i.e. if two or more eigenvalues coincide) the
gauge-invariance is enhanced. It is remarkable that the equation of motion
for the bosonic variable $u_n(x,t)$ is the same as that satisfied by
the fermion field. However, here the boson variables are constrained by
\eighteen.

A detailed analysis of the model along the above lines is in progress and
will appear elsewhere.

\underbar{Acknowledgement}: One of us (S.R.W) would like to thank I. Bakas
for  discussions.

\refout
\end